\documentstyle[12pt]{article}
\begin{document}

\centerline{\large {\bf BFKL pomeron in the next-to-leading approximation}} %
\vskip.3in \centerline{\ V. S. Fadin$^{*}$ and L. N. Lipatov$^{**}$} \vskip%
.2in \centerline{\ $^{*}$Budker Nuclear Physics Institute and} 
\centerline{\ Novosibirsk State University} \centerline{\ Novosibirsk,
630090, Russia} \vskip.2in \centerline{\ $^{**}$Petersburg Nuclear Physics
Institute and} 
\centerline{\ St.Petersburg State  University} \centerline{\ Gatchina,
188350, Russia}

\vskip.6 in \centerline{\bf Abstract} \vskip.1in We find one-loop correction
to the integral kernel of the BFKL equation for the total cross section of
the high energy scattering in QCD and calculate the next-to-leading
contribution to anomalous dimensions of twist-2 operators near $j=1$. \vskip%
.1in

\vskip.2in

The BFKL equation is very important for the theory of the Regge processes at
high energies $\sqrt{s}$ in the perturbative QCD\cite{bfkl}. In particular,
it can be used together with the DGLAP evolution equation \cite{dglap} for
the description of structure functions for the deep inelastic $ep$
scattering at small values of the Bjorken variable $x=-q^2/(2pq)$, where $p$
and $q$ are the momenta of the proton and the virtul photon correspondingly.
But up to recent years the integral kernel for the BFKL equation was known
only in the leading logarithmic approximation (LLA), which did not allow one
to find its region of applicapability, including the scale in transverse
momenta fixing the argument of the QCD coupling constant $\alpha (ck_{\perp
}^2)\,$and the longitudinal scale $\sqrt{s_0}$ for the minimal initial
energy. In this paper we calculate the QCD radiative corrections to this
kernel.

In LLA the gluon is reggeized and the Pomeron is a compound state of two
reggeized gluons. One can neglect multi-gluon components of the Pomeron wave
function also in the next-to-leading logarithmic approximation (NLLA) and
express the total cross-section $\sigma (s)$ for the high energy scattering
of colourless particles $A,B$ in terms of their impact factors $\Phi _i(%
\overrightarrow{q_i})$ and the $t$-channel partial wave $G_\omega (%
\overrightarrow{q},\overrightarrow{q^{\prime }})$ for the reggeized gluon
scattering at $t=0$:

\begin{equation}
\sigma (s)=\int \frac{d^2q}{2\pi q^2}\int \frac{d^2q^{\prime }}{2\pi
q^{\prime 2}}\,\Phi _A(\overrightarrow{q})\,\,\Phi _B(\overrightarrow{%
q^{\prime }})\,\int_{a-i\infty }^{a+i\infty }\frac{d\omega }{2\pi i}\,\left(
\frac s{q\,q^{\prime }}\right) ^\omega \,G_\omega (\overrightarrow{q},%
\overrightarrow{q^{\prime }})\,.  \label{l1}
\end{equation}
Here $\overrightarrow{q}$ and $\overrightarrow{q^{\prime }}$ are transverse
momenta of gluons with the virtualities $-\overrightarrow{q}^2\equiv -q^2$
and $-\overrightarrow{q^{\prime }}^2\equiv -q^{\prime 2}$ correspondingly, $%
s=2p_Ap_B$ is the squared invariant mass of the colliding particles with
momenta $p_A$ and $p_B$. Note, that the dependence of the Regge factor from $%
q$ and $q^{\prime }$ is natural from the point of view of the
Watson-Sommerfeld representation for high energy scattering amplitudes. The change of the energy scale in this
factor leads generally to the corresponding modification of the impact factors and the BFKL 
equation for $G_\omega$ but the physical results are not changed.

Using the dimensional regularization in the $\overline{MS}$-scheme to
renormalize the QCD coupling constant and to remove infrared divergencies in
the intermediate expressions, we write the generalized BFKL equation for $%
G_\omega (\overrightarrow{q},\overrightarrow{q^{\prime }})$ in the following
form

\begin{equation}
\omega \,G_\omega (\overrightarrow{q},\overrightarrow{q^{\prime }})=\delta
^{D-2}(\overrightarrow{q}-\overrightarrow{q^{\prime }})+\int d^{D-2}%
\widetilde{q}\,\,\,K(\overrightarrow{q},\overrightarrow{\widetilde{q}}%
)\,G_\omega (\overrightarrow{\widetilde{q}},\overrightarrow{q^{\prime }})\,.
\label{l2}
\end{equation}
Here 
\begin{equation}
K(\overrightarrow{q_1},\overrightarrow{{q}_2})=2\,\omega (q_1)\,\delta
^{(D-2)}(\overrightarrow{q_1}-\overrightarrow{q_2})+K_r(\overrightarrow{q_1},%
\overrightarrow{q_2})\,,  \label{l3}
\end{equation}
and the space-time dimension is $D=4+2\varepsilon $ for $\epsilon
\rightarrow 0$. The gluon Regge trajectory $\omega (q)$ and the integral
kernel $K_r(\overrightarrow{q_1},\overrightarrow{q_2})$, related with the
real particle production, are expanded in the series over the QCD coupling
constant 
\begin{equation}
\omega (q)=\omega _B(q)+\omega ^{(2)}(q)+...\,,\,\,K_r(\overrightarrow{q_1},%
\overrightarrow{q_2})=K_r^B(\overrightarrow{q_1},\overrightarrow{q_2}%
)+K_r^{(1)}(\overrightarrow{q_1},\overrightarrow{q_2})+...\,\,,\,\,
\label{l4}
\end{equation}
where the Born expressions, corresponding to LLA, are 
\begin{equation}
\omega _B(q)=-\overline{g_\mu ^2}(\frac 2\varepsilon +2\ln \frac{q^2}{\mu ^2}%
)\,,\,\,\,\,K_r^B(\overrightarrow{q_1},\overrightarrow{q_2})=\frac{4\,%
\overline{g_\mu ^2}\,\mu ^{-2\varepsilon }}{\pi ^{1+\varepsilon }\Gamma
(1-\varepsilon )}\,\frac 1{\left( \overrightarrow{q_1}-\overrightarrow{q_2}%
\right) ^2}\,\,.  \label{l5}
\end{equation}
Here 
\[
\overline{g_\mu ^2}=\frac{g_\mu ^2\,N_c\,\Gamma (1-\varepsilon )}{(4\pi
)^{2+\varepsilon }}\, 
\]
for the colour group $SU(N_c)$ and $g_\mu $ is the QCD coupling constant
fixed at the normalization point $\mu $ in the $\overline{MS}$-scheme.

The program of calculating next-to-leading corrections to the BFKL equation
was formulated several years ago \cite{qmr}. It was shown, that the
corrections can be expressed through the Born production amplitudes in the
quasi-multi-regge kinematics with the use of the unitarity conditions in the 
$s$ and $t$ channels. These amplitudes were constructed in terms of various
reggeon-particle vertices. The gauge-invariant action containing all such
vertices was formulated recently \cite{action}.

The two-loop correction $\omega ^{(2)}(q)$ to the gluon Regge trajectory is
known \cite{traj} and for massless quarks can be written as follows

\[
\omega ^{(2)}(q)=-\bar g_\mu ^4\left[ \left( \frac{11}3-\frac 23\frac{n_f}{%
N_c}\right) \left( \frac 1{\varepsilon ^2}-\ln {}^2\left( \frac{\vec q^2}{%
\mu ^2}\right) \right) \right. 
\]
\begin{equation}
\left. +\left( \frac{67}9-\frac{\pi ^2}3-\frac{10}9\frac{n_f}{N_c}\right)
\left( \frac 1\varepsilon +2\ln \left( \frac{\vec q^2}{\mu ^2}\right)
\right) -\frac{404}{27}+2\zeta (3)+\frac{56}{27}\frac{n_f}{N_c}\right] ~.
\label{l6}
\end{equation}
where $n_f$ is the number of light quarks. The poles in $\varepsilon =0$
correspond to infrared divergencies cancelled in the total cross-section.

The one-loop correction to the integral kernel $K_r^{(1)}(\overrightarrow{q_1%
},\overrightarrow{q_2})$ is obtained as a sum of two contributions. The
first one is related with the one-loop virtual correction to the one-gluon
production cross-section \cite{loop} and the second one is determined by the Born cross-sections for 
production of two gluons \cite{gluons} and quark-antiquark pair \cite{jiaf1}. The final result for $K_r^{(1)}(\overrightarrow{q_1},\overrightarrow{q_2})$
can be written as follows

\[
K_r^{(1)}(\overrightarrow{q_1},\overrightarrow{q_2})=\frac{4\overline{g}_\mu
^4\,\mu ^{-2\varepsilon }}{\pi ^{1+\varepsilon }\,\Gamma (1-\varepsilon )}%
\left\{ \frac 1{(\,{\vec q_1}-{\vec q_2)}^2}\left[ \left( \frac{11}3-\frac{%
2n_f}{3N_c}\right) \frac 1\varepsilon \left( 1-\left( \frac{(\overrightarrow{%
q_1}-\overrightarrow{q_2})^2}{\mu ^2}\right) ^\varepsilon (1-\varepsilon ^2%
\frac{\pi ^2}6)\right) \right. \right. \, 
\]
\[
\left. \left. +\left( \frac{(\overrightarrow{q_1}-\overrightarrow{q_2})^2}{%
\mu ^2}\right) ^\varepsilon \left( \frac{67}9-\frac{\pi ^2}3-\frac{10}9\frac{%
n_f}{N_c}+\varepsilon \left( -\frac{404}{27}+14\zeta (3)+\frac{56}{27}\frac{%
n_f}{N_c}\right) \right) \right] \right. 
\]
\[
\left. -\left( 1+\frac{n_f}{N_c^3}\right) \frac{2{\vec q_1}^{~2}{\vec q_2}%
^{~2}-3({\vec q_1}{\vec q_2})^2}{16{\vec q_1}^{~2}{\vec q_2}^{~2}}\left(
\frac 2{{\vec q_2}^{~2}}+\frac 2{{\vec q_1}^{~2}}+(\frac 1{{\vec q_2}%
^{~2}}-\frac 1{{\vec q_1}^{~2}})\ln \frac{{\vec q_1}^{~2}}{{\vec q_2}^{~2}}%
\right) -\frac 1{(\vec q_1-\vec q_2)^2}\left( \ln \frac{{\vec q_1}^{~2}}{{%
\vec q_2}^{~2}}\right) ^2\right. 
\]
\[
\left. +\frac{2(\vec q_1^2-\vec q_2^2)}{(\vec q_1-\vec q_2)^2(\vec q_1+\vec
q_2)^2}\left( \frac 12\ln \left( \frac{{\vec q_1}^{~2}}{{\vec q_2}^{~2}}%
\right) \ln \left( \frac{{\vec q_1}^{~2}{\vec q_2}^{~2}(\vec q_1-\vec q_2)^4%
}{({\vec q_1}^{~2}+{\vec q_2}^{~2})^4}\right) +L\left( -\frac{{\vec q_1}^{~2}%
}{{\vec q_2}^{~2}}\right) -L\left( -\frac{{\vec q_2}^{~2}}{{\vec q_1}^{~2}}%
\right) \right) \right. 
\]
\[
\left. -\left( 3+(1+\frac{n_f}{N_c^3})\left( 1-\frac{({\vec q_1}^{~2}+{\vec
q_2}^{~2})^2}{8{\vec q_1}^{~2}{\vec q_2}^{~2}}-\frac{2{\vec q_1}^{~2}{\vec
q_2}^{~2}-3{\vec q_1}^{~4}-3{\vec q_2}^{~4}}{16{\vec q_1}^{~4}{\vec q_2}^{~4}%
}({\vec q_1}{\vec q_2})^2\right) \right) \int_0^\infty \frac{dx\,\ln \left| 
\frac{1+x}{1-x}\right| }{{\vec q_1}^{~2}+x^2{\vec q_2}^{~2}}\right. 
\]
\begin{equation}
\left. -\left( 1-\frac{(\vec q_1^2-\vec q_2^2)^2}{(\vec q_1-\vec q_2)^2(\vec
q_1+\vec q_2)^2}\right) \left( \int_0^1-\int_1^\infty \right) \frac{dz\,\ln 
\frac{(z{\vec q_1})^2}{({\vec q_2})^2}}{(\vec q_2-z\vec q_1)^2}\right\} \,,
\label{l7}
\end{equation}
where 
\[
L(z)=\int_0^z\frac{dt}t\ln (1-t)\,,\,\,\zeta (n)=\sum_{k=1}^\infty k^{-n}. 
\]

 Note, that the gluonic part of the above expression for $K_r^{(1)}(\overrightarrow{q_1},%
 \overrightarrow{q_2})$ is different from the corresponding piece of the so called "irreducible part" of the kernel, which was constructed by the authors of ref.\cite{jiaf} and it leads to different estimates of radiative corrections to the intercept of the BFKL Pomeron. This difference is due to the fact, that the 
result of Ref.\cite{jiaf}) is incomplete. So-called scale-dependent contributions were not calculated by these authors and in their opinion there is an ambiguity in fixing these contributions.  We stress again, that physical results 
do not depend on the energy scale in the Regge factors in eq.(1) because
its change is compensated by the corresponding modification of the impact factors and the 
kernel. Moreover, this modification does not have any influence on the 
correction to the Pomeron intercept.
 
In our approach the contribution to the total cross section from one gluon
production in the central rapidity region can be presented as follows

\begin{equation}
\sigma _{2\rightarrow 3}=\int \frac{d^2q}{2\pi q^2}\int \frac{d^2q^{\prime }%
}{2\pi q^{\prime 2}}\Phi _A(\overrightarrow{q})\,\Phi _B(\overrightarrow{%
q^{\prime }})\,\int_{q^{\prime }k_{\perp }/s}^{k_{\perp }/q}\frac{d\beta }%
\beta \left( \frac{k_{\perp }\,}{q\,\beta }\right) ^{2\omega
(q)}R^g(q,q^{\prime })\,\left( \frac{s\,\beta \,\,}{q^{\prime }k_{\perp }}%
\right) ^{2\omega (q^{\prime })}\,,  \label{l8}
\end{equation}
where $\overrightarrow{k_{\perp }}=\overrightarrow{q}-\overrightarrow{%
q^{\prime }}$ is the emitted gluon transverse momentum and $\beta $ is its
Sudakov variable, $\beta =(kp_B)/(p_Ap_B)$. Note, that acting more accurately
one should introduce in the impact factors and in the limits of integration
over $\beta $ some intermediate cut-offs in such way, that the physical
results do not depend on them. The quantity $R^g(q,q^{\prime })$ is the
total contribution to the BFKL kernel from one gluon producton and it is
expressed through the product of the Reggeon-Reggeon-gluon vertices $\Gamma
_\mu $ (see \cite{loop}) with a subtraction of the terms proportional to the
product of the gluon Regge trajectories $\omega _B$ and the vertices $\Gamma
_\mu ^B$ in the Born approximation

\[
R^g(q,q^{\prime })\sim \Gamma _\mu \Gamma _\mu -\left( \omega _B(q)\ln \frac{%
\mu ^4}{q^2k_{\perp }^2}+\omega _B(q^{\prime })\ln \frac{\mu ^4}{q^{\prime
2}k_{\perp }^2}\right) \Gamma _\mu ^B\Gamma _\mu ^B\,. 
\]
The neccesity of this subtraction is related with the renormalization of the
Regge factors in eq. (\ref{l8}) in comparison with their definition in ref. 
\cite{loop}.

The ultraviolet divergency in the integral over the relative gluon rapidity
for the contribution to the BFKL kernel from the two gluon emission \cite
{gluons} is cancelled with the corresponding divergency, related with the
infrared cut-off in the relative rapidity for the produced gluons in the
multi-Regge kinematics. The analogous cancellation is implied in the loop
corrections to the impact factors\thinspace $\Phi _i(\overrightarrow{q})$.

The solution of the inhomogenious BFKL equation can be presented as a linear
combination of a complete set of the solutions of a homogenious equation
proportional to spherical harmonics in the $D-2$ dimensional space. The
biggest eigen value $\omega_P$, leading to a rapid increase of total
cross-sections $\sigma (s)\sim s^{\omega_P}$ with energy, corresponds to a
spherically symmetric eigen function. Therefore it is natural to average the
BFKL kernel over the angle between the momenta $\overrightarrow{q_1}$ and $%
\overrightarrow{q_2}$:

\[
\overline{K_r^{(1)}(\overrightarrow{q_1},\overrightarrow{q_2})}=\frac{4%
\overline{g}_\mu ^4\,\mu ^{-2\epsilon }}{\pi ^{1+\epsilon }\,\Gamma
(1-\epsilon )}\left\{ \frac 1{\left| q_2^2-q_1^2\right| }\left[ \left( \frac{%
11}3-\frac{2n_f}{3N_c}\right) \frac 1\epsilon \left( \left( \frac{\left|
q_2^2-q_1^2\right| }{(\max (q_1^2,q_2^2))}\right) ^{2\varepsilon
}(1+\varepsilon ^2\frac{\pi ^2}3)\right. \,\right. \right. 
\]
\[
\left. \left. \left. -\left( \frac{\left( q_2^2-q_1^2\right) ^4}{\mu ^2(\max
(q_1^2,q_2^2))^3}\right) ^\varepsilon \left( 1+\varepsilon ^2\frac{5\pi ^2}%
6\right) \right) +\left( \frac{\left| q_2^2-q_1^2\right| ^4}{\mu ^2(\max
(q_1^2,q_2^2))^3}\right) ^\varepsilon \left( \frac{67}9-\frac{\pi ^2}3-\frac{%
10}9\frac{n_f}{N_c}\right. \right. \right. 
\]
\[
\left. \left. \left. +\varepsilon \left( -\frac{404}{27}+14\zeta (3)+\frac{56%
}{27}\frac{n_f}{N_c}\right) \right) \right] -\frac 1{32}\left( 1+\frac{n_f}{%
N_c^3}\right) \left( \frac 2{{q_2^2}}+\frac 2{{q_1^2}}+(\frac 1{{q_2^2}%
}-\frac 1{{q_1^2}})\ln \frac{{q_1^2}}{{q_2^2}}\right) \right. 
\]
\[
\left. -\frac 1{|q_1^2-q_2^2|}\left( \ln \frac{{q_1^2}}{{q_2^2}}\right)
^2-\left( 3+(1+\frac{n_f}{N^3})\left( \frac 34-\frac{(q_1^2+q_2^2)^2}{%
32q_1^2q_2^2}\right) \right) \int_0^\infty \frac{dx}{q_1^2+x^2q_2^2}\ln
\left| \frac{1+x}{1-x}\right| \right. 
\]
\begin{equation}
\left. +\frac 1{q_2^2+q_1^2}\left( \frac{\pi ^2}3-4L(\min (\frac{q_1^2}{q_2^2%
},\frac{q_2^2}{q_1^2})\right) \right\} \,.  \label{l9}
\end{equation}

Instead of the dimensional regularization we can remove the infrared
divergencies in the kernel by introducing a fictious gluon mass $\lambda $.
Using eqs. (\ref{l4})-(\ref{l6}), it is possible to verify that the averaged
kernel (\ref{l9}) at $\varepsilon \rightarrow 0$ is equivalent to the
expression 
\[
\overline{K(\overrightarrow{q_1},\overrightarrow{q_2})}=-2\frac{\alpha
_s(\mu ^2)N_c}{\pi ^2}\left\{ \ln \frac{q_1^2}{{\lambda }^2}+\frac{\alpha
_s(\mu ^2)N_c}{4\pi }\left[ \left( \frac{11}3-\frac{2n_f}{3N_c}\right)
\left( \ln \frac{{q_1}^2}{{\lambda }^2}\ln \frac{{\mu }^2}{{\lambda }^2}+%
\frac{\pi ^2}{12}\right) \right. \right. 
\]
\[
\left. \left. +\left( \frac{67}9-\frac{\pi ^2}3-\frac{10}9\frac{n_f}{N_c}%
\right) \ln \frac{{q_1}^2}{{\lambda }^2}-3\zeta (3)\right] \right\} \delta
(q_1^2-q_2^2)+\frac{\alpha _s(\mu ^2)N_c}{\pi ^2}\frac{\theta
(|q_1^2-q_2^2|-\lambda ^2)}{|q_1^2-q_2^2|} 
\]
\[
\times \left\{ 1-\frac{\alpha _s(\mu ^2)N_c}{4\pi }\left[ \left( \frac{11}3-%
\frac{2n_f}{3N_c}\right) \ln {\left( \frac{|q_1^{~2}-q_2^{~2}|^2}{\max
(q_1^2,q_2^2)\mu ^2}\right) }-\left( \frac{67}9-\frac{\pi ^2}3-\frac{10}9%
\frac{n_f}{N_c}\right) \right] \right\} 
\]
\[
-\frac{\alpha _s^2(\mu ^2)N_c^2}{4\pi ^3}\left\{ \frac 1{32}\left( 1+\frac{%
n_f}{N_c^3}\right) \left( \frac 2{{q_2}^2}+\frac 2{{q_1}^2}+(\frac 1{{q_2}%
^2}-\frac 1{{q_1}^2})\ln \frac{{q_1}^2}{{q_2}^2}\right) \right. 
\]
\[
\left. +\frac 1{|q_1^2-q_2^2|}\left( \ln \frac{{q_1}^2}{{q_2}^2}\right)
^2+\left( 3+(1+\frac{n_f}{N^3})\left( \frac 34-\frac{(q_1^2+q_2^2)^2}{%
32q_1^2q_2^2}\right) \right) \int_0^\infty \frac{dx}{q_1^2+x^2q_2^2}\ln
\left| \frac{1+x}{1-x}\right| \right. 
\]
\begin{equation}
\left. -\frac 1{q_2^2+q_1^2}\left( \frac{\pi ^2}3-4L(\min (\frac{q_1^2}{q_2^2%
},\frac{q_2^2}{q_1^2}))\right) \right\} \,,  \label{l10}
\end{equation}
defined in the two-dimensional transverse space, with $\lambda \rightarrow 0$%
. Of course, the dependence from $\lambda $ disappears when the kernel acts
on a function. Moreover, the representation (\ref{l10}) permits to find such
form of the kernel for which this cancellation is evident: 
\[
\int d^{D-2}q_2\,\,\,K(\overrightarrow{q_1},\overrightarrow{q_2})\,f(q_2^2)=%
\frac{\alpha _s(\mu ^2)N_c}\pi \int dq_2^2\left\{ \frac
1{|q_1^{~2}-q_2^{~2}|}\left( f(q_2^2)-2\frac{\min (q_1^2,q_2^2)}{%
(q_1^2+q_2^2)}f(q_1^2)\right) \right. 
\]
\[
\left. \times \left[ 1-\frac{\alpha _s(\mu ^2)N_c}{4\pi }\left( \left( \frac{%
11}3-\frac{2n_f}{3N_c}\right) \ln {\left( \frac{|q_1^{~2}-q_2^{~2}|^2}{\max
(q_1^2,q_2^2)\mu ^2}\right) }-\left( \frac{67}9-\frac{\pi ^2}3-\frac{10}9%
\frac{n_f}{N_c}\right) \right) \right] \right. 
\]
\[
\left. -f(q_2^2)\frac{\alpha _s(\mu ^2)N_c}{4\pi }\left[ \frac 1{32}\left( 1+%
\frac{n_f}{N_c^3}\right) \left( \frac 2{{q_2}^2}+\frac 2{{q_1}^2}+(\frac 1{{%
q_2}^2}-\frac 1{{q_1}^2})\ln \frac{{q_1}^2}{{q_2}^2}\right) \right. \right. 
\]
\[
\left. \left. +\frac 1{|q_1^2-q_2^2|}\left( \ln \frac{{q_1}^2}{{q_2}^2}%
\right) ^2+\left( 3+(1+\frac{n_f}{N^3})\left( \frac 34-\frac{(q_1^2+q_2^2)^2%
}{32q_1^2q_2^2}\right) \right) \int_0^\infty \frac{dx}{q_1^2+x^2q_2^2}\ln
\left| \frac{1+x}{1-x}\right| \right. \right. 
\]
\begin{equation}
\left. \left. -\frac 1{q_2^2+q_1^2}\left( \frac{\pi ^2}3-4L(\min (\frac{q_1^2%
}{q_2^2},\frac{q_2^2}{q_1^2}))\right) \right] \right\} +\frac{\alpha
_s^2(\mu ^2)N_c^2}{4\pi ^2}\left( 6\zeta (3)-\frac{5\pi ^2}{12}\left( \frac{%
11}3-\frac{2n_f}{3N_c}\right) \right) f(q_1^2)\,.  \label{l11}
\end{equation}
The $\mu $ - dependence in the right hand side of this equality leads to the
violation of the scale invariance and is related with running the QCD
coupling constant.

The form (\ref{l11}) is very convenient for finding the action of the kernel
on the eigenfunctions $q_2^{2(\gamma -1)}$ of the Born kernel:

\begin{equation}
\int d^{D-2}q_2\,\,\,K(\overrightarrow{q_1},\overrightarrow{q_2})\left( 
\frac{q_2^2}{q_1^2}\right) ^{\gamma -1}=\frac{\alpha _s(q_1^2)\,N_c\,}\pi
\left( \chi (\gamma )+\delta (\gamma )\frac{\alpha _s(q_1^2)N_c}{4\,\pi }%
\right) \,,\, \label{l12}
\end{equation}
were within our accuracy we expressed the result in terms of the running coupling constant
\[
 \alpha _s(q^2)=\frac{\alpha _s(\mu ^2)}{1+\frac{\alpha _s(\mu
 ^2)N_c}{4\pi }%
 \left( \frac{11}3-\frac{2n_f}{3N_c}\right) \ln {\left( \frac{q^2}{\mu
 ^2}%
 \right) }}\simeq \alpha _s(\mu ^2)\left( 1-\frac{\alpha _s(\mu
 ^2)N_c}{4\pi }%
 \left( \frac{11}3-\frac{2n_f}{3N_c}\right) \ln {\left( \frac{q^2}{\mu
 ^2}%
 \right) }\right)\,\, .
 \]
 The introduced quantity $\chi (\gamma ) $ is proportional to the
 eigenvalue of the Born kernel
 \begin{equation}
 \chi (\gamma )=2\psi (1)-\psi (\gamma )-\psi (1-\gamma )\,,\,\,\,\psi
 (\gamma )=\Gamma ^{\prime }(\gamma )/\Gamma (\gamma )\,,  \label{l13}
 \end{equation}
 and  the correction $\delta (\gamma)$ is given below 
\[
\delta (\gamma )=-\left[ \left( \frac{11}3-\frac{2n_f}{3N_c}\right) \frac
12\left( \chi ^2(\gamma )-\psi ^{\prime }(\gamma )+\psi ^{\prime }(1-\gamma
)\right) -\left( \frac{67}9-\frac{\pi ^2}3-\frac{10}9\frac{n_f}{N_c}\right)
\chi (\gamma )\right. 
\]
\[
\left. -6\zeta (3)+\frac{\pi ^2\cos(\pi \gamma )}{\sin^2(\pi \gamma
)(1-2\gamma )}\left( 3+\left( 1+\frac{n_f}{N_c^3}\right) \frac{2+3\gamma
(1-\gamma )}{(3-2\gamma )(1+2\gamma )}\right) \right. 
\]
\begin{equation}
\left. -\psi ^{\prime \prime }(\gamma )-\psi ^{\prime \prime }(1-\gamma )-%
\frac{\pi ^3}{\sin (\pi \gamma )}+4\phi (\gamma )\right] \,.  \label{l12a}
\end{equation}
The function $\phi
(\gamma )$ is 
\[
\phi (\gamma )=-\int_0^1\frac{dx}{1+x}\left( x^{\gamma -1}+x^{-\gamma
}\right) \int_x^1\frac{dt}t\ln (1-t) 
\]
\begin{equation}
=\sum_{n=0}^\infty (-1)^n\left[ \frac{\psi (n+1+\gamma )-\psi (1)}{(n+\gamma
)^2}+\frac{\psi (n+2-\gamma )-\psi (1)}{(n+1-\gamma )^2}\right] \,.
\label{l14}
\end{equation}
Note, that the contribution in the last line of eq. (\ref{l12a}) is different
from the result of ref.\cite{jiaf}, which leads to different estimates of
the next-to-leading corrections to the intercept of the BFKL pomeron, for the reasons mentioned before.

Almost all terms in the right hand side of eq. (\ref{l12}) except the
contribution 
\[
\Delta (\gamma )=\frac{\alpha _s^2(\mu ^2)N_c^2}{4\pi ^2}\left( \frac{11}3-%
\frac{2n_f}{3N_c}\right) \frac 12\left( \psi ^{\prime }(\gamma )-\psi
^{\prime }(1-\gamma )\right) \, 
\]
are symmetric to the transformation $\gamma \leftrightarrow 1-\gamma $.
Moreover, it is possible to cancel $\Delta (\gamma )$ if one would redefine
the function $q^{2(\gamma -1)}$ by including in it the logarithmic factor $%
 \left(\frac{\alpha _s(q^2)}{\alpha _s(\mu^2)}\right)^{-1/2}$.

Because the radiative correction to $\omega $ is negative, it is convenient
to introduce the relative correction $c(\gamma )$ by the definition $\delta (\gamma )=-c(\gamma )\,\chi (\gamma )$. 
In particular, for the symmetric point $\gamma =1/2$, corresponding to the
rightmost singularity of the $t$-channel partial wave, we obtain 
\[
c\left( \frac 12\right) =2\left( \frac{11}3-\frac{2n_f}{3N_c}\right) \ln 2-%
\frac{67}9+\frac{\pi ^2}3+\frac{10}9\frac{n_f}{N_c}+\frac 1{4\ln 2}\left[
16\int_0^1\arctan (\sqrt{t})\ln (\frac 1{1-t})\frac{dt}t\right. 
\]
\begin{equation}
\left. +22\zeta (3)+\frac{\pi ^3}2\left( \frac{27}{16}+\frac{11}{16}\frac{n_f%
}{N_c^3}\right) \right] =25,8388+0.1869\frac{n_f}{N_c}+10.6584\frac{n_f}{%
N_c^3},  \label{l15}
\end{equation}
which is almost two times larger, than its estimate in ref. \cite{jiaf}. For
example, if $\alpha _s(q^2)=0.15$, where the Born intercept is $\omega_P
^B=4N_c\alpha _s(q^2)/\pi \,\ln 2=.39714$, the relative correction for $%
n_f=0 $ is very big: 
\[
\frac {\omega_P}{\omega_P^B}=1-c\left( \frac 12\right) \,\frac{\alpha _s(q^2)}{%
4\pi }N_c=0.0747. 
\]
The maximal value of $\omega_P \simeq 0.1$ is obtained for $\alpha _s(q^2)\simeq 0.08$. These numerical estimates show, that in the kinematical region of HERA
probably it is not enough to take into account only the next-to-leading
correction. The value of this correction strongly depends on its representation. For example, if one 
takes into account the next-to-leading correction by
the corresponding increase of the argument of the running QCD coupling
constant, the intercept of the pomeron turns out to be only two times smaller,
than its Born value.

Due to the effect of running the coupling constant the eigenfunctions of
the NLLA kernel, which can be easily found, do not coincide with $q^{2(\gamma -1)}$; moreover, the
position and the nature of the Pomeron singularity are strongly affected by the nonperturbative effects \cite{conf}. The BFKL equation
with the next-to-leading corrections can be considered as the quantized
version  of the renormalization group equations.  Expression (\ref{l15})
can be used only for a rough estimate of the power of the energy dependence
of the total cross section.

The above results can be applied for the calculation of anomalous dimensions
of the local operators in the vicinity of the point $\omega =J-1=0$. To
begin with, it is necessary to emphasize the difference between the $t$%
-channel partial wave ${\cal G}_\omega ^{gB}(q^2)$ for the reggeized gluon
scattering off the colourless particle $B$ at $t=0$ 
\begin{equation}
{\cal G}_\omega ^{gB}(q^2)=\int \frac{d^2q^{\prime }}{2\pi q^{\prime 2}}%
G_\omega (\overrightarrow{q},\overrightarrow{q^{\prime }})\Phi _B(%
\overrightarrow{q^{\prime }})\,  \label{l17}
\end{equation}
and the deep-inelastic moments ${\cal F}_\omega ^{gB}(q^2)$ defined as
follows 
\begin{equation}
{\cal F}_\omega ^{gB}(q^2)=\int_{q^2}^\infty \frac{ds}s\left( \frac
s{q^2}\right) ^{-\omega }\sigma ^{gB}(q^2,s)\,,  \label{l18}
\end{equation}
where 
\begin{equation}
\sigma ^{gB}(q^2,s)=\int \frac{d^2q^{\prime }}{2\pi q^{\prime 2}}\,\Phi _B(%
\overrightarrow{q^{\prime }})\,\int_{a-i\infty }^{a+i\infty }\frac{d\omega }{%
2\pi i}\,\left( \frac s{q\,q^{\prime }}\right) ^\omega \,G_\omega (%
\overrightarrow{q},\overrightarrow{q^{\prime }})\,.  \label{l19}
\end{equation}
This difference was not essential in LLA but becomes important in NLLA.
Whereas the $t$-channel partial wave${\cal \ G}_\omega ^{gB}(q^2)$ obeys the
integral equation of the type (\ref{l2}) with the same kernel (and the
inhomogenious term equal to $\frac{\Phi _B(\overrightarrow{q})}{%
2\pi q^2}$), the kernel of the corresponding equation for the deep
inelastic moments ${\cal F}_\omega ^{gB}(q^2)$ in NLLA is 
\begin{equation}
{\widetilde{K}}(\overrightarrow{q_1},\overrightarrow{q_2})=K(\overrightarrow{%
q_1},\overrightarrow{q_2})-\frac 12\int d^{D-2}{q^{\prime }}\,\,\,K^B(%
\overrightarrow{q_1},\overrightarrow{{q^{\prime }}})\,\ln {\frac{q^{\prime
}{}^2}{q_1^2}}K^B(\overrightarrow{q^{\prime }},\overrightarrow{{q_2}})\,,
\label{l20}
\end{equation}
where $K^B(\overrightarrow{q^{\prime }},\overrightarrow{{q_2}})$ is the LLA
kernel. The action of the modified kernel on the Born eigen functions $%
q_2^{2(\gamma -1)}$ can be calculated easily:

\begin{equation}
\int d^{D-2}q_2\,\,\,\widetilde{K}(\overrightarrow{q_1},\overrightarrow{q_2}%
)\,\,\left( \frac{q_2^2}{q_1^2}\right) ^{\gamma -1}=\frac{\alpha
_s(q_1^2)\,N_c\,}\pi \left( \chi (\gamma )+\widetilde{\delta }(\gamma )\frac{%
\alpha _s(q_1^2)N_c}{4\,\pi }\right) \,,\, \label{l21}
\end{equation}
where

\begin{equation}
\widetilde{\delta }(\gamma )=\delta (\gamma )-2\chi (\gamma )\,\chi ^{\prime
}(\gamma )\,.  \label{l22}
\end{equation}
The anomalous dimensions 
\[
\gamma =\gamma _0(\alpha _s/\omega )+\alpha _s\gamma _1(\alpha _s/\omega ) 
\]
of the twist-2 operators near point $\omega =0$ are determined from the
solution of the equation

\[
\omega =\frac{\alpha _sN_c}\pi \left( \chi (\gamma )+\widetilde{\delta }%
(\gamma )\frac{\alpha _s N_c}{4\,\pi }\right) =\frac{\alpha _sN_c}\pi
\left( \frac 1\gamma +O(\gamma ^2)\right) -\frac{\alpha _s^2 N_c^2}{%
4\,\pi ^2}\left( \frac{11+2n_f/N_c^3}{3\,\gamma ^2}+\right. 
\]
\begin{equation}
\left. \frac{n_f(10+13/N_c^2)}{9\gamma \,N_c}+\frac{395}{27}-2\zeta (3)-%
\frac{11}3\frac{\pi ^2}6+\frac{n_f}{N_c^3}(\frac{71}{27}-\frac{\pi ^2}%
9)+O(\gamma )\right)   \label{l22b}
\end{equation}
for $\gamma \rightarrow 0$. The singularity in $\delta (\gamma )\simeq
-2/\gamma ^3$ for $\gamma \rightarrow 0$ is exactly cancelled in the
transition to $\widetilde{\delta }(\gamma )$ because $2\chi (\gamma )\,\chi
^{\prime }(\gamma )\,\simeq -2/\gamma ^3$ in the same limit. It gives a
possibility to find the corrections to the anomalous dimensions. In
particular for the low orders of the perturbation theory we reproduce the
known results and predict the higher loop correction for $\omega \rightarrow
0$:
\[
 \gamma =\frac{\alpha _s\,N_c}\pi (\frac 1\omega
 -\frac{11}{12}-\frac{n_f}{%
 6N_c^3}+O(\omega))-\left( \frac{\alpha _s\,}\pi \right)
 ^2\frac{n_f\,N_c}{6\,\omega }%
 (\frac 53+\frac{13}{6N_c^2}+O(\omega))
 \]
 \begin{equation}
 -\frac 1{4\omega ^2}\left( \frac{\alpha _sN_c}\pi \right) ^3
 \left( \frac{395}{27}-2\zeta (3)-%
 \frac{11}3\frac{\pi ^2}6+\frac{n_f}{N_c^3}(\frac{71}{27}-\frac{\pi
 ^2}%
 9)+O(\omega)\right) +O\left(\frac{\alpha _s^4}{\omega^4}\right)\,\,.
\label{l24}
\end{equation}

 Note, that our formulas for the anomalous dimensions are
 different
 from the expressions obtained by authors of ref.\cite{jiaf}. Their results 
are not complete because they did not calculate the contributions of so called energy-scale dependent parts of
 the kernel. 

\vspace{1cm} \noindent

{\large {\bf Acknowledgements}}\\

We want to thank Universit\"at Hamburg and DESY for the hospitality during
our stay in Germany. The fruitful discussions with J. Bartels, M. Ciafaloni
and J. Bl\"umlein were very helpful. The work was supported by the INTAS and
RFFI grants. One of us (L.N.L.) is thankful to Deutsche
Forschunggemeinschaft for the grant, which gave him a possibility to work on
this problem at the Hamburg University.


\vspace{1cm} \noindent

\begin{thebibliography}{9}
\bibitem{bfkl}  V. S. Fadin, E. A. Kuraev, L. N. Lipatov, Phys. Lett. ${\bf {%
B60}}$ (1975) 50;\\Ya. Ya. Balitsky, L. N. Lipatov, Sov. J. Nucl. Phys. $%
{\bf {28}}$ (1978) 822.

\bibitem{dglap}  V. N. Gribov, L. N. Lipatov, Sov. J. Nucl. Phys. ${\bf {15}}
$ (1972) 438;\\L. N. Lipatov, Sov. J. Nucl. Phys. ${\bf {20}}$ (1975) 94;\\%
G. Altarelli, G. Parisi, Nucl. Phys. ${\bf {B26}}$ (1977) 298;\\Yu. L.
Dokshitzer, Sov. Phys. JETP ${\bf {46}}$ (1977) 641.

\bibitem{qmr}  L. N. Lipatov, V. S. Fadin, Sov. J. Nucl. Phys. ${\bf {50}}$
(1989) 712.

\bibitem{action}  L. N. Lipatov, Nucl. Phys. ${\bf {B452}}$ (1995) 369;
Physics Reports ${\bf {286}}$ (1997) 132.

\bibitem{traj}  V. S. Fadin, R. Fiore, M. I. Kotsky, Phys. Lett. ${\bf {B359}%
}$ (1995) 181; ${\bf {B387}}$ (1996) 593.

\bibitem{loop}  V. S. Fadin, L. N. Lipatov, Nucl. Phys. ${\bf {B406}}$
(1993) 259;\\V. S. Fadin, R. Fiore and A. Quartarolo, Phys. Rev. ${\bf {D50}}
$ (1994) 5893;\\V. S. Fadin, R. Fiore, M. I. Kotsky, Phys. Lett. ${\bf {B389}%
}$ (1996) 737.

\bibitem{gluons}  V. S. Fadin, L. N. Lipatov, Nucl.Phys. ${\bf {B477}}$
(1996) 767;\\V. S. Fadin, M. I. Kotsky, L. N. Lipatov, Phys. Lett. ${\bf {%
B415}}$ (1997) 97

\bibitem{jiaf1}  S. Catani, M. Ciafaloni and F.Hautman, Phys. Lett. ${\bf {B242}}$ (1990) 97; Nucl.Phys. ${\bf {B366}}$ (1991) 135;  \\G. Camici and  M. Ciafaloni, Phys. Lett. ${\bf {B386}}$ (1996) 341; Nucl.Phys. ${\bf {B496}}$ (1997) 305; \\ 
V. S. Fadin, R. Fiore, A. Flashi, M. I. Kotsky,
BUDKERINP-97-86, hep-ph 9711427. 

\bibitem{jiaf}  G. Camici, M. Ciafaloni, Phys. Lett. ${\bf {B412}}$ (1997)
396.

\bibitem{conf}  L. N. Lipatov, JETP ${\bf {63}}$ (1986) 904; \\G. Camici and M. Ciafaloni,  Phys. Lett. ${\bf {B395}}$ (1997) 118.
\end{thebibliography}
\end{document}